\documentclass[pra, twocolumn, showpacs, showkeys, preprintnumbers]{revtex4}
\usepackage{setspace}
\usepackage{graphicx}
\usepackage{amsmath}
\usepackage{amssymb}
\usepackage{latexsym}

\usepackage{natbib}
\begin{document}
\title{Disruption of reflecting Bose-Einstein condensates due to inter-atomic interactions and quantum noise}

\author{R.G. Scott}   
\author{D.A.W. Hutchinson}
\author{C.W. Gardiner}
\affiliation{The Jack Dodd Centre for Photonics and Ultra-Cold Atoms, Department of Physics, University of Otago, P.O. Box 56, Dunedin, New Zealand.}

\date{\today}

\begin{abstract}
We perform fully three-dimensional simulations, using the truncated Wigner method, to investigate the reflection of Bose-Einstein condensates from abrupt potential barriers. We show that the inter-atomic interactions can disrupt the internal structure of a cigar-shaped cloud with a high atom density at low approach velocities, damping the center-of-mass motion and generating vortices. Furthermore, by incorporating quantum noise we show that scattering halos form at high approach velocities, causing an associated condensate depletion. We compare our results to recent experimental observations.
\end{abstract}


\maketitle





\section{INTRODUCTION}
Recent experiments~\cite{tom,tom2} have demonstrated quantum reflection of Bose-Einstein condensates (BECs) from a silicon surface, and showed the potential applications of semiconductor surfaces as atom mirrors and traps. This is very exciting in the light of the intense interest in precise manipulation of cold atoms with atom chips~\cite{folman,hansel}, which have been used to reflect, transport, split, and recombine atom clouds~\cite{hinds,cassettari,hansel2,reichel,dekker,lau}. In principle, semiconductor surfaces have many advantages over conventional atom chips: they can be precisely patterned, require no magnetic fields or currents, and hence avoid the problems of fragmentation~\cite{esteve}. In order to optomize the effective manipulation of atoms, theoretical work is required to understand the reflection process, and in particular, how it may disrupt the cloud. Previous theoretical work~\cite{me}, based on solving the cylindrically symmetric Gross-Pitaevskii equation (GPE), showed that inter-atomic interactions can disrupt the internal structure of the cloud on reflection, creating topological excitations such as solitons and vortices. Recent theoretical work on colliding BECs~\cite{adam,adamlong}, however, demonstrated that the bare GPE is insufficient for describing the production of scattering halos and the associated depletion of the BEC, developing the truncated Wigner method (TWM)~\cite{adam,adamlong,steel,sinatra}, which includes quantum noise, as an alternative method to model this process. Furthermore, the experimental results have confirmed that large scattering halos are indeed created for particular parameter regimes~\cite{tom2}.

In this paper, we reproduce the experimental results by performing fully three-dimensional simulations using the TWM. In contrast to the bare GPE, this method models quantum fluctuations of the condensate, and can hence describe processes which require spontaneous initiation. One such process is the pairwise scattering of condensate atoms into unoccupied modes, which is the precursor for the formation of a scattering halo. Once these previously unoccupied modes have some finite occupation, further atoms enter them via Bosonic stimulation, creating a scattering halo. By comparing multiple simulations the TWM may also be used to calculate the condensate depletion associated with the halo.  

We explore the role of the cloud geometry, density and approach speed in the reflection of BECs from abrupt potential steps. We find that the disruption due to the inter-atomic interactions, which we refer to as {\it interferential disruption}, is most pronounced for cigar-shaped BECs with high atom densities, approaching the potential barrier along its long axis at low velocities. This effect generates vortices and causes an associated damping of the center-of-mass motion. The production of scattering halos is also enhanced by elongating the cloud and increasing the density, but in contrast to the interferential disruption, is most pronounced at high approach velocity. As mentioned previously, the creation of the scattering halo causes a depletion of the condensate. We compare our theoretical results to recent experimental observations of excited reflected clouds and scattering halos~\cite{tom,tom2}.

We study two sets of $^{23}$Na BEC parameters, referred to as {\it A} and {\it B}, taken from recent experiments on quantum reflection of BECs from silicon surfaces. BEC {\it B} is more than twice as dense (its equilibrium peak density $n_{0}$ is $5.2 \times 10^{12}$ cm$^{-3}$) as BEC {\it A} ($n_{0}=2.0\times 10^{12}$ cm$^{-3}$). Both BECs have different widths in each coordinate direction. The long axis of BEC {\it B} is always perpendicular to the barrier, but we examine the dynamics of BEC {\it A} in three orientations, in which the cloud is accelerated in each of the three principal coordinate directions. By studying the behaviour of a single BEC in different orientations we draw conclusions about the role of BEC shape and orientation in determining the subsequent dynamics. We show that our theoretical predictions for the two sets of parameters are in good agreement with the experimental observations. 

The paper is organised as follows: in section~\ref{theory} we discuss the theoretical model used to simulate the BEC dynamics, in section~\ref{A} we present results for BEC {\it A}, in section~\ref{B} we present results for BEC {\it B}, and in section~\ref{conclusions} we summarize our findings and conclude.

\section{THEORETICAL MODEL OF THE BEC DYNAMICS AND QUANTUM FLUCTUATIONS OF THE FIELD}
\label{theory}
\subsection{The truncated Wigner method}
Heuristically, the TWM can be thought of as a classical field technique which simulates quantum vacuum fluctuations by adding appropriate classical random fluctuations to the coherent field of the BEC's intial state. The added fluctuations are referred to as {\it virtual particles}, and the atoms in the BEC initial state are referred to as the {\it real particles}. This approach is valid for {\it high densities} of real particles. Specifically, in a homogenous system, the number of real particles must be much greater than the number of virtual particles. The derivation of this method is outlined in Ref.~\cite{adam} and described in detail in Ref.~\cite{adamlong}. The dynamical equations of the TWM are identical to the projected GPE~\cite{mat}.   

The wavefunction $\Psi\left({\bf x},t\right)$ is modelled by the mode expansion
\begin{equation}
\Psi\left({\bf x},t\right)=\frac{1}{\sqrt{V}}\sum_{j=1}^{j=M} \alpha_{j}\left(t\right)e^{i{\bf k}_{j}\cdot{\bf x}},
\label{eq:mode}
\end{equation} 
where $t$ is time, $V = L_{x}L_{y}L_{z}$ is the volume contained in the coordinate space and $\alpha_{j}\left(t\right)$ is the amplitude of the mode with wavevector ${\bf k}_{j}$, normalised such that $\sum_{j=1}^{j=M} \alpha_{j}^{*}\left(t\right)\alpha_{j}\left(t\right)$ is the total number of atoms $N$. The mode space is spherical, being contained within a maximum cut-off wavevector modulus to prevent Fourier aliasing~\cite{adamthesis}. The wavevectors are defined as  
\begin{equation}
{\bf k}_{j} = \frac{2\pi p_{j}}{L_{x}} {\bf \hat{k}}_{x} + \frac{2\pi q_{j}}{L_{y}} {\bf \hat{k}}_{y} + \frac{2\pi r_{j}}{L_{z}} {\bf \hat{k}}_{z}, 
\label{eq:ks}
\end{equation} 
where $p_{j}$, $q_{j}$ and $r_{j}$ are integers. 

For $t<0$, the BEC is held in a magnetic trap of frequencies $\omega_{x}$, $\omega_{y}$ and $\omega_{z}$, centered at $\left(-\Delta x,0,0\right)$, with potential profile 
\begin{equation}
U_{\textrm{trap}} = \frac{m}{2}\left[\omega_{x}^{2}\left(x+\Delta x\right)^{2}+\omega_{y}^{2}y^{2}+\omega_{z}^{2}z^{2}\right],
\end{equation}
where $m$ is the mass of a single $^{23}$Na atom. The initial state is determined by solving the three-dimensional GPE
\begin{equation}
i\hbar\frac{\partial \psi}{\partial t} = \left[-\frac{\hbar^{2}\nabla^{2}}{2m} + U_{\textrm{trap}} + U_{0}\left|\psi\right|^{2}\right] \psi
\label{eq:GPE}
\end{equation}
where 
\begin{equation}
U_{0} = \frac{4\pi\hbar^{2}a}{m},
\end{equation}
in which $a = 2.9$ nm is the s-wave scattering length, using an imaginary time algorithm~\cite{imag}. The quantum fluctuations are introduced by combining this real particle field $\psi\left({\bf x}\right)$ with a field of virtual particles $\chi\left({\bf x}\right)$ to create the total field $\Psi\left({\bf x},0\right)$. The virtual particle field is defined as
\begin{equation}
\chi\left({\bf x}\right)=\frac{1}{\sqrt{V}}\sum_{j=1}^{j=M} \chi_{j}e^{i{\bf k}_{j}\cdot{\bf x}},
\end{equation}
in which the complex amplitudes $\chi_{j}$ have a Gaussian distribution with the properties $\langle\chi_{i}^{*}\chi_{j}\rangle=\frac{1}{2}\delta_{ij}$ and $\langle\chi_{i}\chi_{j}\rangle=0$~\cite{adam,adamlong}. This means that on average each mode is populated by half a virtual particle, so that the total number of virtual particles is approximately $M/2$.

The derivation in Ref.~\cite{adamlong} yields the stochastic differential equation for each mode within the low energy subspace
\begin{equation}
i\hbar\frac{d\alpha_{j}}{dt}=\frac{\hbar^{2}k_{j}^{2}}{2m}\alpha_{j} + \frac{1}{\sqrt{V}}\int e^{-i{\bf k}_{j}\cdot{\bf x}} \left[U_{\textrm{ext}}+U_{0}\left|\Psi\right|^{2}\right] \Psi d{\bf x},
\label{eq:sde}
\end{equation} 
where $U_{\textrm{ext}}$ is the total external potential. We compute the dynamics of the BEC by solving these equations using {\it the Fourth-order Runge-Kutta in the Interaction Picture algorithm} (RK4IP)~\cite{rob}.

\subsection{Application to the reflection problem}
At $t=0$ the magnetic trap is displaced by a distance $\Delta x$ along the $x$-direction, such that it is now centered at $\left(0,0,0\right)$, creating the new potential profile $U_{\textrm{trap}}^{'}$. This accelerates the BEC towards an abrupt potential step $U_{\textrm{wall}}$ of height $V$, which is positive, in the $y-z$ plane at $x=0$, so that 
\begin{equation}
U_{\textrm{wall}} = \left\{ \begin{array}{ll}
V & x \geq 0 \\
0 & x<0
\end{array} \right. .
\end{equation} 
Hence, the impact velocity of the cloud at the wall $v_{x} \approx \omega_{x}\Delta x$~\cite{foot}. If the mean kinetic energy of the atoms at the barrier $\langle E \rangle \approx \frac{m v_{x}^{2}}{2} \ll V$, all atoms are reflected. If $\langle E \rangle \gtrsim V$, there is finite transmission.

In this paper we compare our theoretical predictions to experimental obvservations of quantum reflection from the Casimir-Polder potential~\cite{tom,tom2}. Unfortunately, it is impossible to model the Casimir-Polder potential directly as this accelerates atoms to large wavevectors which exceed our maximum cut-off wavevector, which is limited by computer memory. However, since the Casimir-Polder potential, and particularly that of a pillared surface, varies over a distance of less than the healing length~\cite{tom2}, it can be approximated to an abrupt potential barrier. Furthermore, previous theoretical work~\cite{me} has demonstrated that the BEC dynamics are qualitatively independent of the form of the reflecting potential barrier. 
 
\subsubsection{Absorption of transmitted real particles}
Transmitted atoms are absorbed by an imaginary potential, given by
\begin{equation}
U_{\textrm{imag}} = \left\{ \begin{array}{ll}
-Cx & x>0 \\
0 & x<0
\end{array} \right. ,
\end{equation} 
where $C$ is a positive constant. The imaginary potential introduces a damping term into the equations for the mode amplitudes, which removes atoms for $x>0$. However, the imaginary potential must {\it only} absorb the real particles, {\it not} the virtual particles, hence maintaining the appropriate amount of quantum noise. We overcome this problem by calculating the damping rate of the imaginary potential, and adding extra virtual particles by increasing each mode amplitude $\alpha_{j}$ by
\begin{equation}
\Delta\alpha_{j} = \frac{\Delta V}{\sqrt{\hbar V}} \sum_{\nu=1}^{\nu=\nu_{\textrm{max}}} W_{\nu}e^{i{\bf k}_{j}\cdot{\bf x_{\nu}}} ,
\end{equation}
every time step $\Delta t$~\cite{crispin}, where $\Delta V$ is the volume around one spatial grid point, and the complex spatial process $W_{\nu}$ has the properties $\langle W_{\nu}\rangle=0$, $\langle W_{\nu}W_{\mu}\rangle=0$, and $\langle W_{\nu}^{*}W_{\mu}\rangle=\frac{-\Delta t U_{\textrm{imag}}}{\Delta V}\delta_{ij}$.

Hence the total external potential is $U_{\textrm{ext}} = U_{\textrm{trap}}^{'} + U_{\textrm{wall}} + U_{\textrm{imag}}$.

\section{DYNAMICS OF BEC {\it A}}
\label{A}
\begin{figure}[tbp]
\centering
\includegraphics[keepaspectratio=true, scale=0.6]{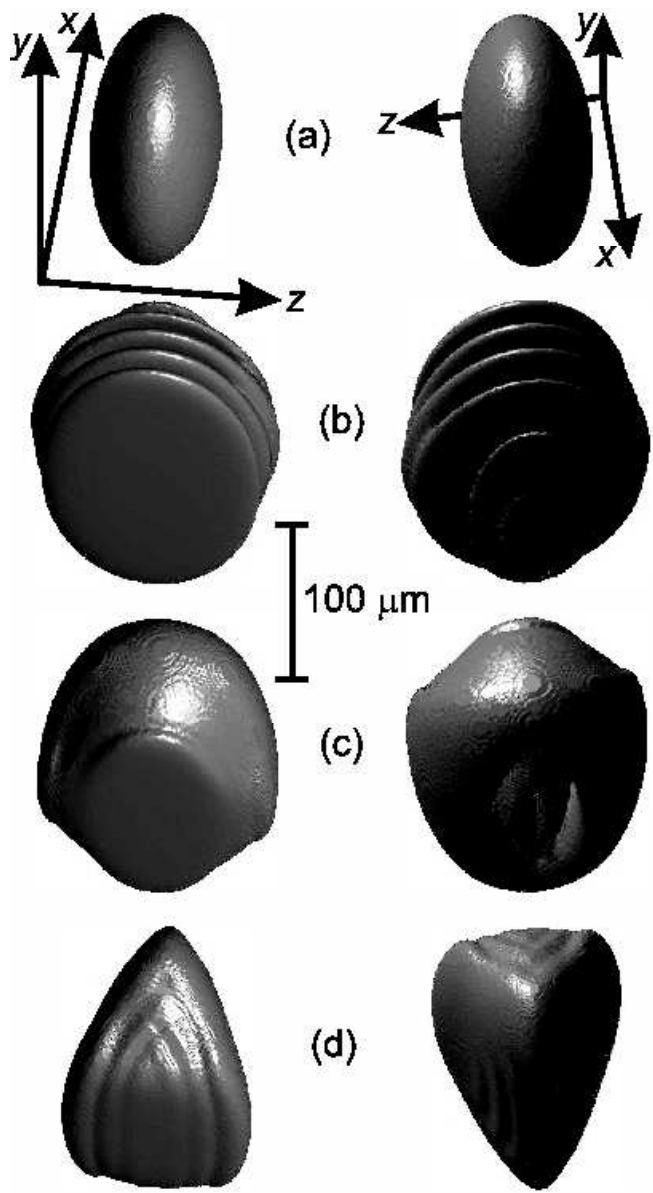}
\caption{Constant density surfaces of BEC {\it A} without quantum fluctuations in orientation 1, reflecting from a barrier of height $V = 10^{-30}$ J for $v_{x}=1.2$ mm s$^{-1}$ at $t=0$ (a), 75 (b), 100 (c) and 125 ms (d). The left and right-hand columns shows views from and towards the barrier respectively [axes are shown in (a)]. The vertical bar shows the scale.}
\label{f1}
\end{figure}

For BEC {\it A}, $N=3 \times 10^5$, $\omega_{x} = 2\pi \times 3.3$ rad s$^{-1}$, $\omega_{y} = 2\pi \times 2.5$ rad s$^{-1}$, and $\omega_{z} = 2\pi \times 6.5$ rad s$^{-1}$. For these parameters $n_{0}=2.0 \times 10^{12}$ cm$^{-3}$. Figure~\ref{f1}(a) shows two views of a constant density surface of the BEC initial state at $t=0$ without added quantum fluctuations (the axes are included in the figure). In this section $V = 10^{-30}$ J, which far exceeds $\frac{m v_{x}^{2}}{2}$ for all $v_{x}$ considered, and hence all atoms are reflected.

Since the cloud is not symmetric, we can explore the role of the BEC geometry by accelerating it in different directions. The parameters as defined above will be referred to as orientation 1. We will later rotate the cloud such that $\omega_{x} = 2\pi \times 2.5$ rad s$^{-1}$, $\omega_{y} = 2\pi \times 3.3$ rad s$^{-1}$, and $\omega_{z} = 2\pi \times 6.5$ rad s$^{-1}$, hence presenting its long axis towards the potential barrier. This will be referred to as orientation 2. Finally, we will rotate the cloud such that $\omega_{x} = 2\pi \times 6.5$ rad s$^{-1}$, $\omega_{y} = 2\pi \times 2.5$ rad s$^{-1}$, and $\omega_{z} = 2\pi \times 3.3$ rad s$^{-1}$, hence presenting its short axis towards the potential barrier. This will be referred to as orientation 3. 


\subsection{Interferential disruption}
\label{mod}
\begin{figure}[tbp]
\centering
\includegraphics[keepaspectratio=true, scale=0.65]{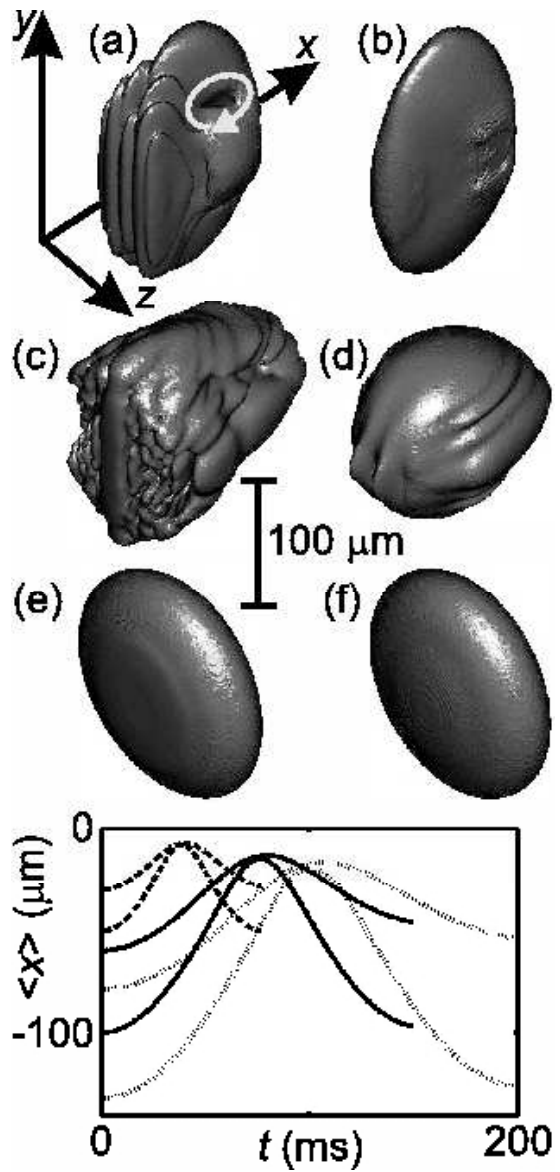}
\caption{Constant density surfaces of BEC {\it A} without quantum fluctuations in orientation 1 [(a) and (b)], orientation 2 [(c) and (d)], and orientation 3 [(e) and (f)] after one complete oscillation, having reflected from a barrier of height $V = 10^{-30}$ J. The left and right-hand figures show results for $v_{x}=1.2$ and $2.1$ mm s$^{-1}$ respectively. The circular arrow in Fig.(a) indicates the direction of circulation around a vortex line. The vertical bar shows the scale, the axes are shown in (a). (g): $\langle x \rangle$ versus $t$ for orientation 1 (solid curves), orientation 2 (dotted curves) and orientation 3 (dashed curves). In all cases the upper and lower curves correspond to $v_{x}=1.2$ and $2.1$ mm s$^{-1}$ respectively.}
\label{f2}
\end{figure}

We initially perform simulations without added quantum noise in order to study the interferential disruption. Figure~\ref{f1} shows constant density surfaces for BEC {\it A} in orientation 1 without quantum fluctuations undergoing reflection from the potential barrier at $t=75$ (b), 100 (c), and 125 ms (d), for $v_{x}=1.2$ mms$^{-1}$. The left and right-hand columns show views of the cloud from and towards the wall respectively [see axes in Fig.~\ref{f1}(a)]. Figure~\ref{f1}(b) shows the mid-point of the oscillation, at which time the cloud is modulated by the standing wave created by the superposition of the incident and reflected matter waves. The left-hand figure shows the flat constant density surface against the potential barrier. Previous theoretical work~\cite{me} demonstrated that the inter-atomic interactions in the high-density peaks of the standing wave cause the momentum of some atoms to be transferred into the radial direction. These atoms appeared as jet-like ``lobes''. In three dimensions, the ``lobes'' can be seen to form more of a ``doughnut'' [Fig.~\ref{f1}(c)]. The trap causes the ``doughnut'' to collapse as it rebounds away from the wall, disrupting the internal structure of the cloud [Fig.~\ref{f1}(d)]. At the end of the oscillation ($t=150$ ms $\approx \pi/\omega_x$), the radial momentum of the ``doughnut'' has created vortex lines which pierce the BEC, roughly parallel to the $y$-axis [Fig.~\ref{f2}(a)]. The circular arrow in Fig.~\ref{f2}(a) shows the direction of quantized circulation. These theoretical findings are in agreement with experimental observations of ``excited and sometimes fragmented'' reflected clouds~\cite{tom}. 

The excitation of the internal structure of the cloud removes energy from the longitudinal motion, and hence this effect is associated with a damping of the center-of-mass motion. The upper solid line in Fig.~\ref{f2}(g) is the mean BEC $x$ position $\langle x \rangle$ as a function of $t$ for these parameters, and it is clearly strongly damped. The effect is described as {\it interferential disruption}, since it arises from an interference pattern, in this case produced by the superposition of the incident and reflected matter waves.

The ``lobes'' form if atoms can move radially outwards from the center of the BEC to its extremities before the reflection process is complete~\cite{me}. The time taken for this to occur has the lower limit $t_{l} = l_{r}/v_{s}$, where $l_{r}$ is the radial half-width of the cloud and $v_{s} \approx \sqrt{h^{2}n_{0}a/\pi m^{2}}$ is the average speed of sound, if the atoms are to travel through the cloud without exciting the BEC. The reflection time is $t_{R}\approx l_{x}/v_{x}$, where $l_{x}$ is the BEC's longitudinal half-width. Consequently we expect to observe lobes and subsequent disruption if 
\begin{equation}
v_{x} \lesssim v_{s} \left(l_{x}/l_{r}\right) .
\label{condition}
\end{equation}

For the simulation described above, we have the interesting situation that equation~\ref{condition} is satisfied if $l_{r}$ is taken to be $l_{z}$, but {\it not} if $l_{r}$ is taken to be $l_{y}$. Consequently, the ``doughnut'' is particularly pronounced in the $z$-direction, and hence its momentum is largest in the $z$-direction. This explains why the resulting vortex lines are roughly parallel to the $y$-axis.  

As $v_{x}$ increases equation~\ref{condition} is no longer satisfied. Figure~\ref{f2}(b) shows the reflected cloud in orientation 1 at $t=150$ ms for $v_{x}=2.1$ mm s$^{-1}$. For these parameters the ``doughnut'' does not form. Hence, at the end of the oscillation the cloud contains no topological excitations and its appearance is similar to the initial state at $t=0$. Consequently, there is little damping of the center-of-mass motion, as shown by the lower solid curve in Fig.~\ref{f2}(g). 

We may control the disruption by changing the geometry of the cloud, i.e. by changing the ratio $l_{x}/l_{r}$ in equation~\ref{condition}. The ratio $l_{x}/l_{r}$ is largest for orientation 2, and smallest for orientation 3, so we would expect more and less disruption for these parameters respectively. This is supported by our numerical simulations. The BEC in orientation 2 is highly disrupted for $v_{x}=1.2$ mm s$^{-1}$ and contains many topological excitations [Fig.~\ref{f2}(c)], and is mildly disrupted for $v_{x}=2.1$ mm s$^{-1}$ [Fig.~\ref{f2}(d)]. However, the BEC in orientation 3 shows virtually no disruption at either impact velocity [Fig.~\ref{f2}(e) and (f)]. Again we find that severe disruption of the BEC internal structure is associated with strong damping of the center-of-mass motion: the damping is largest for orientation 2 [dotted curves in Fig.~\ref{f2}(g)], and smallest for orientation 3 [dashed curves in Fig.~\ref{f2}(g)]. These results illustrate that we expect severe interferential disruption for large $l_{x}$ and small $l_{r}$, i.e. a cigar-shaped BEC with its long axis perpendicular to the barrier, as predicted by equation~\ref{condition}. Conversely, we expect mild interferential disruption for small $l_{x}$ and large $l_{r}$, i.e. a pancake-shaped BEC with its short axis perpendicular to the barrier.

\subsection{Inclusion of quantum fluctuations}
\begin{figure}[tbp]
\centering
\includegraphics[keepaspectratio=true, scale=0.65]{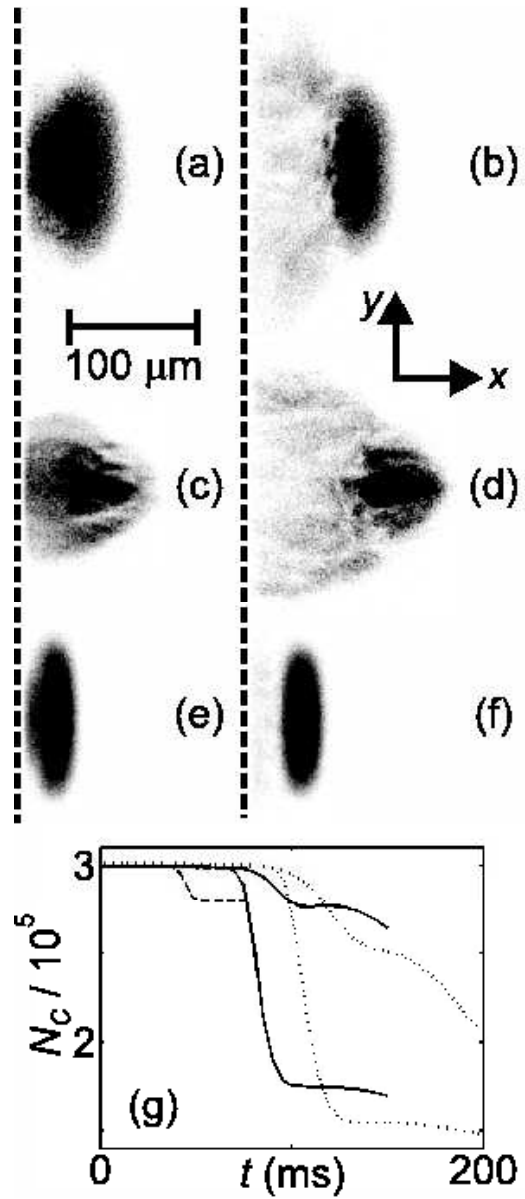}
\caption{Simulated absorption images of BEC {\it A} including quantum fluctuations in orientation 1 [(a) and (b)], orientation 2 [(c) and (d)], and orientation 3 [(e) and (f)] after one complete oscillation, having reflected from a barrier of height $V = 10^{-30}$ J. The left and right-hand figures show results for $v_{x}=1.2$ and $2.1$ mm s$^{-1}$ respectively. The horizontal bar shows the scale, and the dashed vertical lines indicate the position of the barrier. (g): $N_{C}$ versus $t$ for orientation 1 (solid curves), orientation 2 (dotted curves) and orientation 3 (dashed curves). In all cases the upper and lower curves correspond to $v_{x}=1.2$ and $2.1$ mm s$^{-1}$ respectively.}
\label{f3}
\end{figure}

We also investigated the role of quantum fluctuations of the field by adding noise to the initial state at $t=0$, as described in section~\ref{theory}. Figure~\ref{f3} shows the resulting simulated absorption images of the BEC in the $y-x$ plane after one complete oscillation for various orientations and approach velocities. The interferential disruption has least effect in this plane for the reasons explained in section~\ref{mod}. For the parameters taken from the first BEC quantum reflection experiments~\cite{tom} (orientation 1), the scattering halo is very weak for an approach velocity of 1.2 mm s$^{-1}$ [Fig.~\ref{f3}(a)]. This is unsurprising as scattering halos were not observed in the experiments. For the larger approach velocity of 2.1 mm s$^{-1}$ [Fig.~\ref{f3}(b)] the halo is much clearer, but still relatively faint. This result is confirmed in Fig.~\ref{f3}(g), which shows the coherent number of atoms $N_{C}$ as function of $t$ for $v_{x} = 1.2$ and 2.1 mm s$^{-1}$ (upper and lower solid curves respectively), calculated by the method described in Ref.~\cite{adamlong}. As the BEC reflects, coherent atoms from the BEC are transferred into the incoherent scattering halo, depleting the condensate. The final condensate fraction is much less (57\% compared to 88\%) for the larger $v_{x}$ of 2.1 mm s$^{-1}$ than for $v_{x}=1.2$ mm s$^{-1}$. 

If the BEC is orientated such that its long axis is perpendicular to the wall (orientation 2), the scattering halo is much more pronounced [Fig.~\ref{f3}(c) and (d)], and there is a greater associated depletion of the condensate [dotted lines in Fig.~\ref{f3}(g)]. Conversely, the halo is only just visible for the BEC in orientation 3 [Fig.~\ref{f3}(e) and (f)]. These results indicate that the production of scattering halos is most pronounced for cigar-shaped BECs, approaching the potential barrier along its long axis with large $v_{x}$. 

\section{DYNAMICS OF BEC {\it B}}
\label{B}
\begin{figure}[tbp]
\centering
\includegraphics[keepaspectratio=true, scale=1.0]{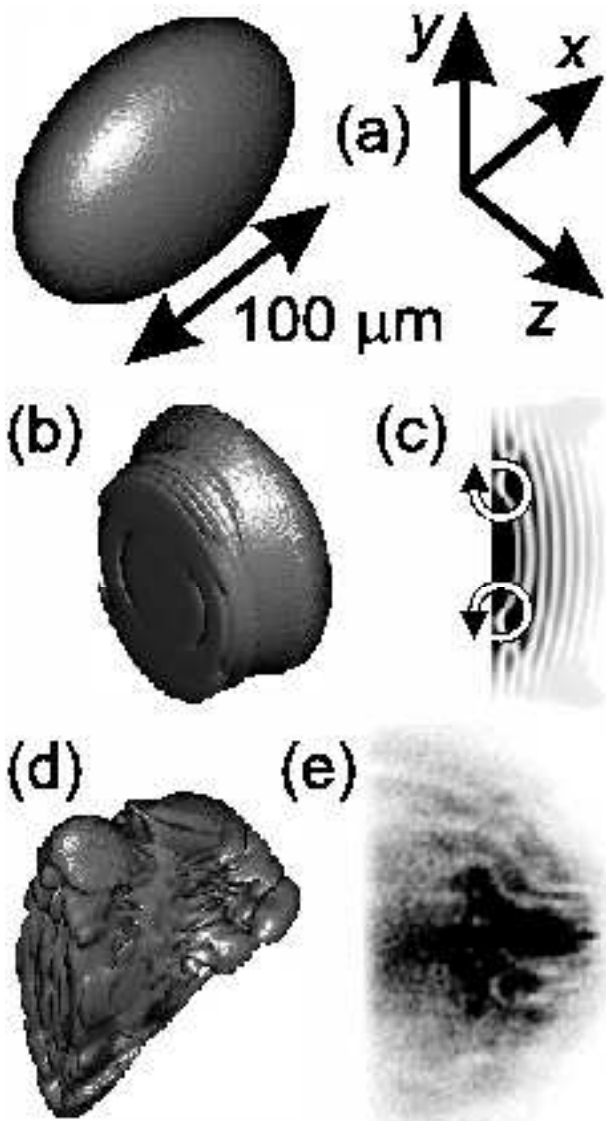}
\caption{Constant density surfaces of BEC {\it B} without quantum fluctuations reflecting from a barrier of height $V=7.97 \times 10^{-32}$ J for $v_{x}=2.1$ mm s$^{-1}$ at $t=0$ (a), 70 (b) and 120 ms (d). The axes are shown in (a). (c): Slice through the BEC in the $y-x$ plane at $t = 70$ ms. The arrows show the direction of circulation around a vortex ring. (e): Simulated absorption image of BEC {\it B} including quantum fluctuations in the $y-x$ plane at $t=120$ ms. The arrow in Fig.(a) indicates the scale.}
\label{f4}
\end{figure}

For BEC {\it B}, $N=10^6$, $\omega_{x} = 2\pi \times 4.2$ rad s$^{-1}$, $\omega_{y} = 2\pi \times 5.0$ rad s$^{-1}$, and $\omega_{z} = 2\pi \times 8.2$ rad s$^{-1}$. These parameters give $n_{0}=5.2 \times 10^{12}$ cm$^{-3}$, which is more than double that in BEC {\it A}. Figure~\ref{f4}(a) shows a constant density surface of BEC {\it B} without added quantum fluctuations at $t=0$. In this section, $V$ is reduced below $\frac{m v_{x}^{2}}{2}$, in order to investigate the effect of finite transmission. Transmitted atoms are absorbed by an imaginary potential, as described in section~\ref{theory}. 

\subsection{Low approach velocity}
Initially, we set $V=7.97 \times 10^{-32}$ J and $v_{x}=2.1$ mms$^{-1}$, for which we would expect a reflection probabiliy of 0.5 for a non-interacting plane wave. 

Since the atom density is higher in BEC {\it B}, and its long axis is perpendicular to the barrier, equation~\ref{condition} predicts more pronounced interferential disruption. This is confirmed by the numerical simulations. Figures~\ref{f4}(b) and (d) show constant density surfaces of BEC {\it B} without quantum fluctuations undergoing reflection from the potential barrier at $t=70$ and 120 ms respectively. At $t=70$ ms the standing wave has formed, but the inter-atomic interactions have already caused disruption of the internal structure and ``doughnut'' formation. In this high-density regime, some vortices are formed directly from the modulation of the standing wave, as has been demonstrated in optical lattices~\cite{me2}. The slice through the $y-x$ plane in Fig.~\ref{f4}(c) shows the bending of the nodal lines, which is characteristic of the snake instability~\cite{anderson,dutton}, before the ``doughnut'' has collapsed. A vortex ring has formed where one nodal line is bending and breaking, and is encircled by two arrows, indicating the direction of quantized circulation. At the end of the oscillation [Fig.~\ref{f4}(d)], the internal structure of the cloud is highly disrupted, and contains many topological exciations. This is in agreement with the results of Ref.~\cite{tom2}, which reports the observation of disruption to the BEC internal structure for approach velocities of approximately 2 mm s$^{-1}$ and below. 

A corresponding simulation including quantum fluctuations shows that the higher density also causes more pronounced halo formation. Figure~\ref{f4}(e) is a simulated absorbtion image of the cloud in the $y-x$ plane at $t=120$ ms, showing a very clear, large and dense halo. 

We obtain qualitatively similar results in the case of zero transmission, by setting $V = 10^{-30}$ J, as in section~\ref{A}.

\subsection{High approach velocity and comparison with experiment}
Finally, we increase $V$ to $1.67 \times 10^{-31}$ and $v_x$ to 3 mm s$^{-1}$ in order to compare our predictions directly with experimental results~\cite{tom2}. For these parameters the plane-wave reflection probability is 0.5, approximately the same as that observed in experiment at this $v_{x}$ on reflection from the Casimir-Polder potential of a pillared silicon surface. 

\begin{figure}[tbp]
\centering
\includegraphics[trim=50 0 0 0, keepaspectratio=true, scale=0.6]{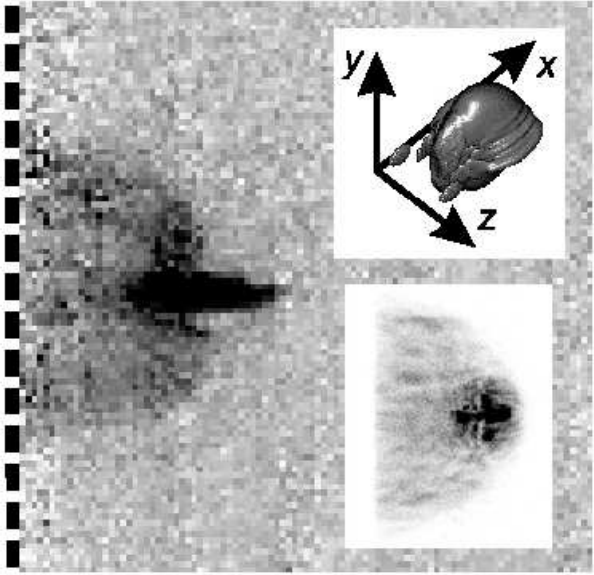} 
\caption{Experimental absorption image of BEC {\it B} for $v_{x}=3.0$ mm s$^{-1}$ at $t=120$ ms, having reflected from the Casimir-Polder potential of a pillared silicon surface. The field of view is 500 $\mu$m, the vertical dashed line indicates the position of the barrier. Lower inset: corresponding simulated absorption image in the $y-x$ plane including quantum fluctuations for reflection from a barrier of height $V=1.67 \times 10^{-31}$ J. Upper inset: equivalent constant density surface excluding quantum fluctuations, axes are shown in the figure.}
\label{f5}
\end{figure}

At this larger velocity there is very little interferential disruption, so without quantum fluctuations the reflected cloud has a reasonably smooth density profile (Fig.~\ref{f5}, upper inset). However, when we include quantum fluctuations an even larger scattering halo is produced, as shown in Fig.~\ref{f5}, lower inset. As before, we obtain qualitatively similar results in the case of zero transmission for $V = 10^{-30}$ J. 

Our simulations agree with the results of Ref.~\cite{tom2}, which reports the observation of large scattering halos for an approach velocity of 3 mm s$^{-1}$, but no interferential disruption for $v_{x} \gtrsim 2$  mms$^{-1}$. An experimental image is shown in Fig.~\ref{f5} for comparison with our simulated absorption image. The experimental and theoretical images both show a dense and coherent cloud that has reflected cleanly from the barrier, superimposed on the background of the incoherent and comparitively dilute halo. These findings demonstrate that the production of scattering halos is most pronounced for dense BECs with large $v_{x}$.

\section{CONCLUSIONS}
\label{conclusions}
It has already been demonstrated that abrupt potential barriers, such as the Casimir-Polder potential of a silicon surface, could be exploited in applications as atom mirrors and traps~\cite{tom}. Furthermore, surface potentials can be routinely patterned in order to create devices to manipulate BECs in more imaginative ways. The results of this paper are more generally relevant to any reflection process, for example using magnetic mirrors~\cite{hinds}, sheets of laser light, or arrays of current-carrying wires~\cite{lau}. These techniques can also be extended to create more exotic potentials, such as diffraction gratings~\cite{gunter,aspect}.

Our findings suggest that these experiments may be most successful for dilute, pancake-shaped BECs, with their short axis perpendicular to the barrier. Furthermore, our results show that there are two competing effects which may spoil the smooth profiles of the incident BECs and produce excitations: interferential disruption and the production of scattering halos. The former is most pronounced at low approach velocity, the latter at high approach velocity. Applications of BEC reflection may be most effective for moderate approach velocities which are above the threshold for interferential disruption, but below the threshold for the formation of a scattering halo. Any future devices based on reflection must also account for possible damping of the center-of-mass oscillations associated with the production of excitations. 

We thank T. Pasquini for helpful discussions, and his kind permission to include the experimental image. This work was the supported by the Royal Society (London), and the Marsden Fund under contracts UOO-0590 and UOO-323.



\begin{thebibliography}{77}	
\bibitem{tom} T.A. Pasquini {\it et al.}, Phys. Rev. Lett. {\bf 93}, 223201 (2004).
\bibitem{tom2} T.A. Pasquini {\it et al.}, Phys. Rev. Lett. {\bf 97}, 093201 (2006).
\bibitem{folman} R. Folman {\it et al.}, Phys. Rev. Lett. {\bf 84}, 4749 (2000).
\bibitem{hansel} W. H\"ansel, P. Hommelhoff, T.W. H\"ansch, J. Reichel, Nature {\bf 413}, 498 (2001).
\bibitem{hinds} E.A. Hinds, I.G. Hughes, J. Phys. D {\bf 32}, R119 (1999). 
\bibitem{cassettari} D. Cassettari, B. Hessmo, R. Folman, T. Maier, J. Schmiedmayer, Phys. Rev. Lett. {\bf 85}, 5483 (2000).
\bibitem{hansel2} W. H\"ansel, J. Reichel, P. Hommelhoff, T.W. H\"ansch, Phys. Rev. Lett. {\bf 86}, 608 (2001).
\bibitem{reichel} J. Reichel, Appl. Phys. B {\bf 75}, 469 (2002).
\bibitem{dekker} N.H. Dekker {\it et al.}, Phys. Rev. Lett. {\bf 84}, 1124 (2000).
\bibitem{lau} D.C. Lau {\it et al.}, Eur. Phys. J. D {\bf 5}, 193 (1999).
\bibitem{esteve} J. Est\'eve {\it et al.}, Phys. Rev. A {\bf 70}, 043629 (2004).
\bibitem{me} R.G. Scott, A.M. Martin, T.M. Fromhold, F.W. Sheard, Phys. Rev. Lett. {\bf 95}, 073201 (2005).
\bibitem{adam} A.A. Norrie, R.J. Ballagh, C.W. Gardiner, Phys. Rev. Lett. {\bf 94}, 040401 (2005).
\bibitem{adamlong} A.A. Norrie, R.J. Ballagh, C.W. Gardiner, Phys. Rev. A {\bf 73}, 043617 (2006).
\bibitem{steel} M.J. Steel {\it et al.}, Phys. Rev. A {\bf 58}, 4824 (1998).
\bibitem{sinatra} A. Sinatra, C. Lobo, Y. Castin, Phys. Rev. Lett. {\bf 87}, 210404 (2001).
\bibitem{mat} M.J. Davis, R.J. Ballagh, K. Burnett, J. Phys. B {\bf 34}, 4487 (2001).
\bibitem{adamthesis} A.A. Norrie, PhD. thesis (2005).
\bibitem{imag} M.L. Chiofalo, S. Succi, M.P. Tosi, Phys. Rev. E {\bf 62}, 7438 (2000).
\bibitem{rob} R.J. Ballagh, Computational methods for nonlinear partial differential equations, {\it http:// www.physics.otago.ac.nz/ research/ uca/ resources/ comp\_lectures\_ballagh.html}, (2000).
\bibitem{foot} Due to the finite spatial width of the BEC there is a spread of incident velocities at the barrier. However, it is possible to define a characteristic velocity as the spread of velocities is symmetric about $v_{x}$, and the oscillation period is independent of $\Delta x$.
\bibitem{crispin} C.W. Gardiner, P. Zoller, {\it Quantum noise}, 2nd edition, Springer-Verlag, (2000).
\bibitem{me2} R.G. Scott {\it et al.}, Phys. Rev. Lett. {\bf 90}, 110404 (2003).
\bibitem{anderson} B.P. Anderson {\it et al.}, Phys. Rev. Lett. {\bf 86}, 2926 (2003).
\bibitem{dutton} Z. Dutton {\it et al.}, Science {\bf 293}, 663 (2001).
\bibitem{gunter} A. G\"unter {\it et al.}, Phys. Rev. Lett. {\bf 95}, 170405 (2005).
\bibitem{aspect} J. Est\'eve {\it et al.},  J. Opt. B {\bf 5}, S103 (2003).

\end{thebibliography}

\end{document}